\documentclass[aps,twocolumn,pra]{revtex4}
\usepackage{epsfig,rotating}
\usepackage{amsmath,amssymb,amsthm}
\usepackage{graphicx}
\usepackage{psfrag}
\usepackage{bm}
\usepackage[dvips]{color}

\begin{document}

\title{Husimi distribution and phase-space analysis of a vibron-model quantum phase transition} 

\author{M. Calixto}
\affiliation{Departamento de Matem\'atica Aplicada, Universidad de Granada,
Fuentenueva s/n, 18071 Granada, Spain}
\author{R. del Real}
\affiliation{Departamento de F\'{\i}sica At\'omica, Molecular y Nuclear, Universidad de Granada, Fuentenueva s/n, 18071 Granada,
Spain}
\author{E. Romera}
\affiliation{Departamento de F\'{\i}sica At\'omica, Molecular y Nuclear and
Instituto Carlos I de F{\'\i}sica Te\'orica y
Computacional, Universidad de Granada, Fuentenueva s/n, 18071 Granada,
Spain}

\date{\today}
\begin{abstract}
The Husimi distribution is proposed for a phase space analysis of quantum phase transitions in the two-dimensional 
$U(3)$ vibron model for $N$-size molecules. We show that the  
inverse participation ratio and Wehrl's entropy of the Husimi distribution 
give sharp signatures of the quantum (shape) phase transition from linear to bent. Numerical results are 
complemented with a variational approach using parity-symmetry-adapted $U(3)$ coherent states, which reach  
the minimum Wehrl entropy $\frac{N(3+2N)}{(N+1)(N+2)}$, in the rigidly linear phase, according to a 
generalized Wehrl-Lieb conjecture. We also propose a characterization  of 
  the vibron-model  quantum phase transition 
by means of the zeros of the Husimi distribution.

\end{abstract}
\maketitle

\section{Introduction}

Quantum Phase Transitions (QPTs) have become an important subject in quantum many
body problems \cite{sachdev}. Unlike classical phase transitions, QPTs take place at 
absolute zero of temperature. Generally speaking, 
one finds different quantum phases connected to specific geometric configurations of
the ground state and related to distinct dynamic symmetries of the Hamiltonian. The  QPT occurs
as a function of a control parameter $\xi$ that appears in the Hamiltonian $H$. For us it will appear  in
the form of a convex combination $H(\xi)=(1-\xi)H_1+\xi H_2$. At $\xi=0$ the system
is in phase I, characterized by the dynamical symmetry $G_1$ of $H_1$, and at $\xi=1$
the system is in phase II, characterized by the dynamical symmetry $G_2$ of
$H_2$. At some critical point $\xi_c\in (0,1)$ there is an abrupt change in the
symmetry and structure of the ground state wavefunction.  This is the case of the so-called `vibron models' 
(see e.g. \cite{Iachello,Iachello1,Iachello2,curro}), interacting boson models which
exhibits a second order shape phase transition from linear to bent.  These models have been  used to study
the rovibrational properties in diatomic and polyatomic molecules 
 and  have turned to be very useful to study symmetry properties of
quantum systems. Other interesting model exhibiting a QPT is the case of an ensemble of atoms interacting with a single bosonic field
mode described by the Dicke Hamiltonian \cite{Dicke}, which shares some features with the present vibron model.

In Quantum Mechanics we have to our disposal several distributions to characterize
phase-space properties \cite{Gerry2005}. On the one hand, the Wigner function, widely used in Quantum Optics. On the other hand, the 
Husimi distribution,  which  is given by the
overlap between a minimal uncertainty (coherent) state and the wavefunction. 
This distribution proves sometimes more convenient because, unlike Wigner distribution, 
it is non-negative. Husimi distribution has been found useful for a
phase-space visualization of a metal-insulator transition \cite{aulbach}, to analyze
quantum chaos in atomic physics \cite{monteiro} or to analyze models in condensed
matter physics \cite{weinmann99}. In addition, we would like to point out that the zeros of the Husimi distribution 
have essential information, in particular, the quantum state can be  described by its distribution of
zeros \cite{Korsch}. They are 
the least probable points in phase space and they have also been considered as a
quantum indicator of classical and quantum chaos \cite{4,dando}.

The Husimi distribution has a great amount of information and it can be useful
to consider informational measures as the so-called inverse participation ratio and Wehrl entropy
\cite{Mintert}. An analysis of the Dicke-model QPT by means of information measures has been done in position and momentum
spaces, separately \cite{pla11,jsm11,epl2012,physica12,renyipra} and through
the Husimi distribution, its marginals {and its participation
  ratio and Wehrl entropy \cite{husimidicke}. Morever, QPT has been characterized
 by means of the zeros of the Husimi distribution in the Dicke model \cite{husimidicke}.}
Here we shall offer an informational description of the vibron-model QPT in 
phase space in terms of the inverse participation
ratio  (and higher moments) and the Wehrl entropy of the Husimi distribution. Additionally we shall 
investigate the visualization of the vibron-model QPT through the zeros of the Husimi
distribution.
 
This article is organized as follows. In Section \ref{sec1} we briefly remind
the {vibron-model  Hamiltonian, introduce coherent states and the Husimi distribution of 
the ground state,  and present the inverse participation ratio and the Wehrl enropy
for the Husimi distribution in a numerical framework.}
 In Section \ref{variationalsec} we study  a variational approximation to the ground state wave function 
in terms of symmetry-adapted coherent states and analyze the information
measures for this approximation in the thermodynamic limit. Zeros of the Husimi (ansatz) distribution are also 
computed  in order to characterize the QPT. 

\section{Vibron model and Husimi distribution}\label{sec1}

 2D-vibron models  describe
a system containing a dipole degree of freedom constrained to planar motion.
Elementary excitations are (creation and annihilation) 2D vector $\tau$-bosons $\{\tau_x^\dag, \tau_y^\dag, \tau_x,
\tau_y\}$ and a scalar $\sigma$-boson $\{\sigma^\dag,\sigma\}$. It is convenient to introduce circular bosons:
$\tau_\pm=\mp(\tau_x\mp i\tau_y)/\sqrt{2}$. The nine generators of the $U(3)$ algebra are bilinear products of
creation and annihilation operators, in particular:
\begin{eqnarray}
 &\hat{n}=\tau_+^\dag\tau_++\tau_-^\dag\tau_-,\, \hat{n}_s=\sigma^\dag\sigma,&\nonumber\\
&\hat{l}=\tau_+^\dag\tau_+-\tau_-^\dag\tau_-,&\\
&\hat{D}_+=\sqrt{2}(\tau^\dag_+\sigma-\sigma^\dag\tau_-),\; \hat{D}_-=\sqrt{2}(-\tau^\dag_-\sigma+\sigma^\dag\tau_+),&\nonumber
\end{eqnarray}
denote the number operator of vector $\hat n$ and scalar $\hat{n}_s$ bosons, 2D angular momentum $\hat{l}$ and dipole $\hat D_\pm$ operators, respectively
(see \cite{curro} for the reminder four operators $\hat{Q}_\pm,\hat{R}_\pm$, which will not be used here).  Assuming
the total number of bosons $\hat{N}=\hat{n}+\hat{n}_\sigma$ and the 2D angular momentum $\hat{l}$ to be conserved,
there are only two dynamical symmetry limits, $G_1=U(2)$ and $G_2=SO(3)$, associated with two algebraic
chains starting from $U(3)$ and ending in $SO(2)$:  the so-called `cylindrical' and `displaced' oscillator chains.  A general Hamiltonian of the $U(3)$ vibron model
with only one- and two-body interactions can be expressed in terms of linear and quadratic Casimir operators of all the
subalgebras contained in the dynamical symmetry algebra chains. To capture the essentials of the phase transition
from the $G_1$-phase (linear) to the $G_2$-phase (bent) it is enough to consider a convex combination of
the linear $C_1(U(2))=\hat{n}$ and quadratic $C_2(SO(3))=\hat{W}^2=(\hat{D}_+\hat{D}_-+\hat{D}_-\hat{D}_+)/2+\hat{l}^2$
Casimir operators of the corresponding dynamical symmetries. In particular, we shall consider the essential
Hamiltonian \cite{curro}
\begin{equation}
 \hat{H}=(1-\xi)\hat{n}+\xi\frac{N(N+1)-\hat{W}^2}{N-1},\label{hamiltonian}
\end{equation}
where the (constant) quantum number $N$ is the total number of bound states that labels the totally symmetric
$(N+1)(N+2)/2$ dimensional representation $[N]$ of $U(3)$. It is known (see \cite{curro} and later on Sec.  \ref{variationalsec}) 
that this model exhibits a (shape) QPT at $\xi_c=0.2$ and we shall see that Wehrl entropies provide sharp indicators of this QPT.

The  Hilbert space is spanned by the orthonormal basis vectors
\begin{equation}
 |N;n,l\rangle=\frac{(\sigma^\dag)^{N-n}(\tau^\dag_+)^{\frac{n+l}{2}}(\tau^\dag_-)^{\frac{n-l}{2}}}
{\sqrt{(N-n)!\left(\frac{n+l}{2}\right)!\left(\frac{n-l}{2}\right)!}}|0\rangle,\label{basis}
\end{equation}
where the bending quantum number $n=N,N-1,N-2,\dots,0$ and the angular momentum
$l=\pm n,\pm(n-2),\dots,\pm 1$ or $0$ ($n=$odd or even) are the eigenvalues of
$\hat{n}$ and $\hat{l}$, respectively. The matrix elements of $\hat{W}^2$ can be easily derived
(see e.g. \cite{curro}):
\begin{equation}
\begin{array}{l}
 \langle N;n',l|\hat{W}^2|N;n,l\rangle= \\  ((N-n)(n+2)+(N-n+1)n+l^2)\delta_{n',n}\\
 -((N-n+2)(N-n+1)(n+l)(n-l))^{\frac{1}{2}} \delta_{n',n-2}\\
 -((N-n)(N-n-1)(n+l+2)(n-l+2))^{\frac{1}{2}}\delta_{n',n+2}.
\end{array}\nonumber
\end{equation}
From these matrix elements, it is easy to see that time evolution preserves the parity $e^{i\pi n}$ of a given state
$|N;n,l\rangle$. That is, the parity operator $\hat\Pi=e^{i\pi \hat{n}}$ commutes with $\hat{H}$ and both operators
can then be  jointly diagonalized. We shall take this fact into account when proposing parity-symmetry-adapted ansatzes 
in subsection \ref{variationalsec}.

\subsection{$SU(3)$ coherent states and Husimi distribution}

Let us use the notation $(a_0,a_1,a_2)\equiv(\sigma,\tau_+,\tau_-)$ for our three oscillator operators. $SU(3)$ projective coherent 
states (CSs) are defined as (for a given $N$)
\begin{eqnarray}
|z_1,z_2\rangle&\equiv&\frac{(a_0^\dag+z_1a_1^\dag+z_2 a_2^\dag)^N|0\rangle}{N! (1+|z_1|^2+|z_2|^2)^{N/2}}\nonumber\\
&=&\sum_{n=0}^N\sum_{m=0}^n\varphi_{n,m}^{(N)}(z_1,z_2)|N;n,l=n-2m\rangle,
\label{cohs}
\end{eqnarray}
with $z_1,z_2\in \mathbb C$ and 
\begin{equation}
 \varphi_{n,m}^{(N)}(z_1,z_2)\equiv\frac{(N!/((N-n)!(n-m)!m!))^{1/2}}{ (1+|z_1|^2+|z_2|^2)^{N/2}} z_1^{n-m}z_2^m\,.
\end{equation}
They can be seen as a generalization of $SU(2)$ spin-$j$ coherent states:
\begin{equation}
|z\rangle={(1+|z|^2)^{-j}}\sum_{m=-j}^j \binom{2j}{j+m}^{1/2}z^{j+m}|j,m\rangle,\label{spinjcs}
\end{equation}
in terms of angular momentum or Dicke states $|j,m\rangle$.

Although $SU(3)$ projective CSs are not an orthonormal set since
\begin{equation}
 \langle z_1,z_2|z'_1,z'_2\rangle=\frac{(1+\bar z_1z'_1+\bar z_2z'_2)^N}{ (1+|z_1|^2+|z_2|^2)^{N/2} (1+|z'_1|^2+|z'_2|^2)^{N/2}},\label{csoverlap}
\end{equation}
they form an overcomplete set 
of the corresponding Hilbert space and fulfill 
the closure relation or resolution of the identity  (see e.g. \cite{Perelomov}):
\begin{equation}
1= \int_{{\mathbb R}^4}|z_1,z_2\rangle\langle z_1,z_2|d\mu(z_1,z_2),\label{closure}
\end{equation}
with 
\begin{equation}
 d\mu(z_1,z_2)=\frac{(N+1)(N+2)}{\pi^2}\frac{d^2 z_1d^2z_2}{(1+|z_1|^2+|z_2|^2)^3}
\end{equation}
the measure on the complex projective (quotient) space $\mathbb CP^2=U(3)/U(1)^3$ and 
$d^2z_{1,2}\equiv d\mathrm{Re}(z_{1,2})d\mathrm{Im}(z_{1,2})$ the usual Lebesgue measure on $\mathbb R^2$ or $\mathbb C$. 
In general, the exact ground state vector $\psi$ will be given as an expansion 
\begin{equation}
 |\psi^{(N)}_\xi\rangle=\sum_{n=0}^{N}\sum_{m=0}^{n} c_{nm}^{(N)}(\xi)|N;n,l=n-2m\rangle,\label{gs}
\end{equation}
where the coefficients $c_{nm}^{(N)}(\xi)$ are calculated by numerical diagonalization of \eqref{hamiltonian}. One can realize that the ground state 
$|\psi^{(N)}_\xi\rangle$ has even-parity since, for instance, $c_{nm}^{(N)}(\xi)=0$ for $n$ odd.

The Husimi distribution ${\Psi}^{(N)}_\xi(z_1,z_2)$ of $\psi^{(N)}_\xi$ is, by definition, given by the squared modulus of the overlap between $|\psi^{(N)}_\xi\rangle$ 
and an arbitrary  coherent state $|z_1,z_2\rangle$, that is:
\begin{eqnarray}
 {\Psi}^{(N)}_\xi(z_1,z_2)&=&|\langle z_1,z_2|\psi^{(N)}_\xi\rangle|^2 \nonumber\\ &=&\sum_{n,n'=0}^{N}\sum_{m,m'=0}^{n} c_{nm}^{(N)}(\xi)
\bar c_{n'm'}^{(N)}(\xi)\nonumber\\ &&\times \varphi_{n,m}^{(N)}(z_1,z_2)\varphi_{n',m'}^{(N)}(\bar z_1,\bar z_2)  \label{husiz}
\end{eqnarray}
and normalized according to:
\begin{equation}
 \int_{\mathbb R^4}  {\Psi}^{(N)}_\xi(z_1,z_2) d\mu(z_1,z_2)=1.\label{Qnorm}
\end{equation}
This can be seen as an alternative (coherent state) representation to the usual position $q$ and momentum $p$ representations traditionally given by 
$\psi(q)=\langle q|\psi\rangle$ and $\tilde\psi(p)=\langle p|\psi\rangle$, respectively (see \cite{nuestro} for the expression of  $|\psi^{(N)}_\xi\rangle$ in position 
representation in terms of Hermite polynomials).

\subsection{Moments and R\'enyi-Wehrl entropy of the Husimi distribution}

Important quantities to visualize the QPT in the vibron model across the critical point $\xi_c$ will be the 
$\nu$-th moments of the Husimi distribution \eqref{husiz}:
\begin{equation}
 M_{N,\nu}(\xi)= \int_{{\mathbb R}^4}({\Psi}^{(N)}_\xi(z_1,z_2))^\nu  d\mu(z_1,z_2)
 .\label{momentsnu}
\end{equation}
Note that $M_{N,1}=1$ since $\Psi^{(N)}_\xi$ is normalized \eqref{Qnorm}. Among all moments we shall single-out 
the so-called ``inverse participation ratio'' (IPR)  $P_N(\xi)=M_{N,2}(\xi)$ which somehow measures the (de-)localization 
of $\Psi^{(N)}_\xi$ across the phase transition.  The `classical' (versus quantum von Neumann) R\'enyi-Wehrl entropy is then defined as:
\begin{equation}
W_{N,\nu}(\xi)=\frac{1}{1-\nu}\ln(M_{N,\nu}(\xi)),\label{wehrlnu}
\end{equation}
for $\nu\not=1$. For $\nu=1$ we have the usual Wehrl entropy
\begin{equation}
W_{N}(\xi)=-\int_{{\mathbb R}^4} {\Psi}^{(N)}_\xi(z_1,z_2)\ln ({\Psi}^{(N)}_\xi(z_1,z_2)) d\mu(z_1,z_2)
\,.\label{wehrl1}
\end{equation}

\subsection{Numerical Results}

We have solved the vibron-model numerically, calculating the coefficients $c_{nm}^{(N)}(\xi)$ in \eqref{gs} by
numerical diagonalization of \eqref{hamiltonian}.

\begin{figure}
\includegraphics[width=4cm]{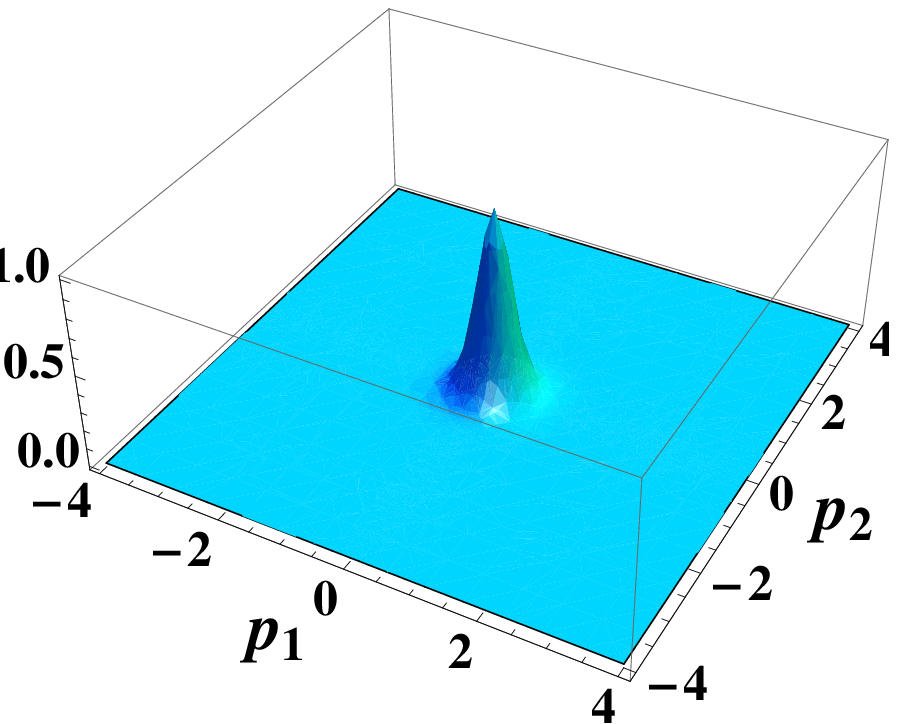}
\includegraphics[width=4cm]{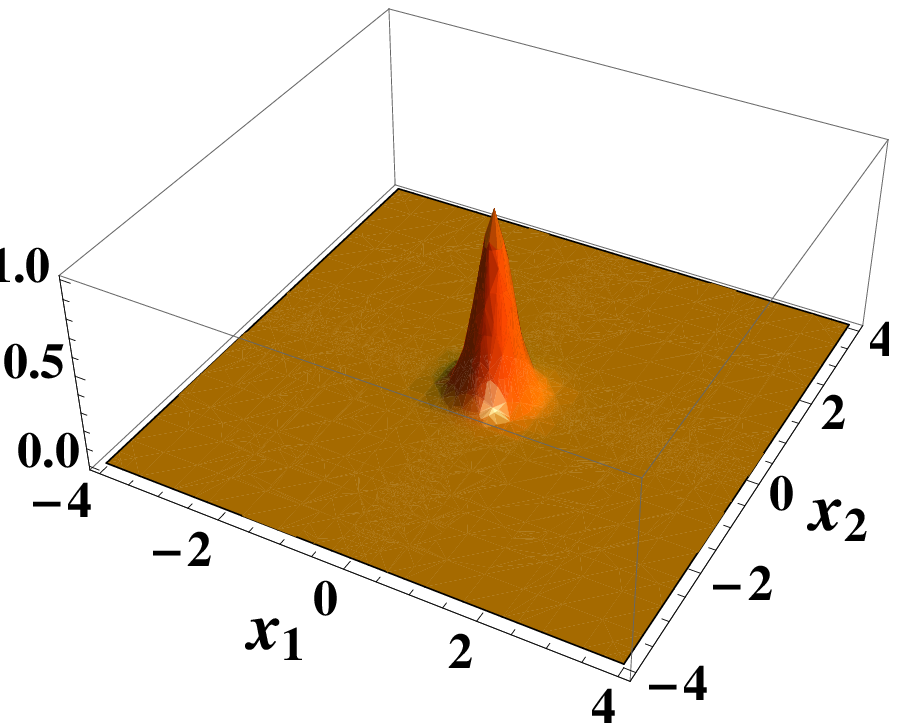}
\includegraphics[width=4cm]{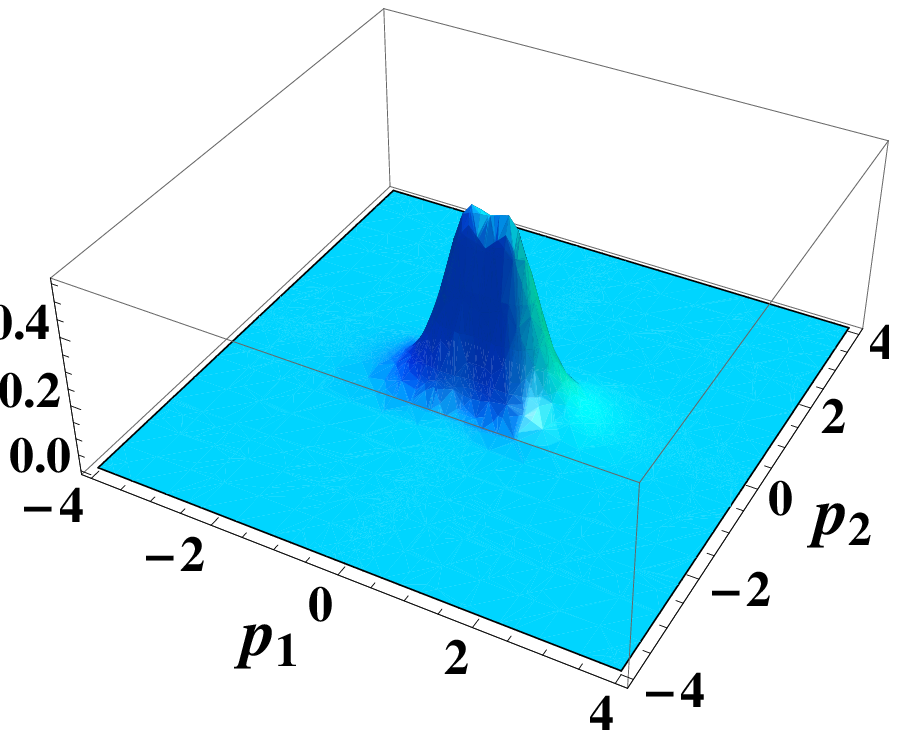}
\includegraphics[width=4cm]{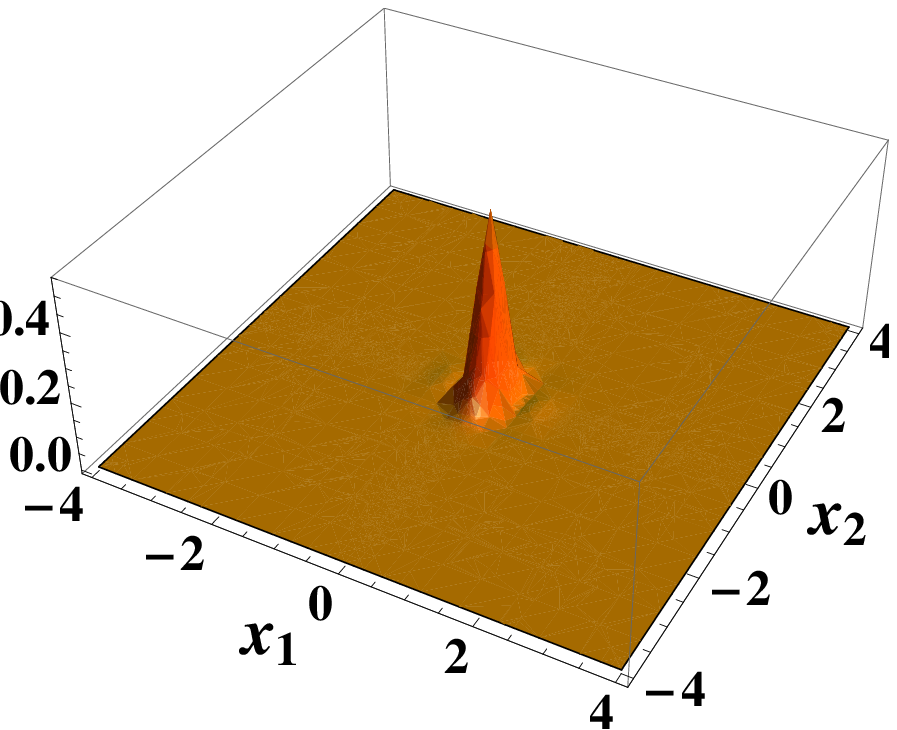}
\includegraphics[width=4cm]{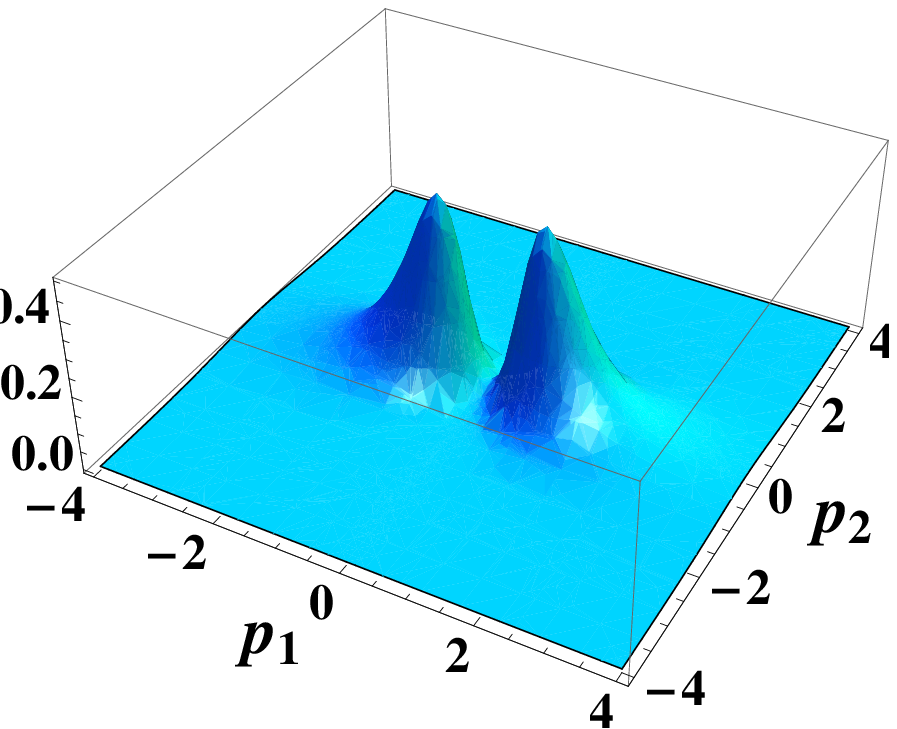}
\includegraphics[width=4cm]{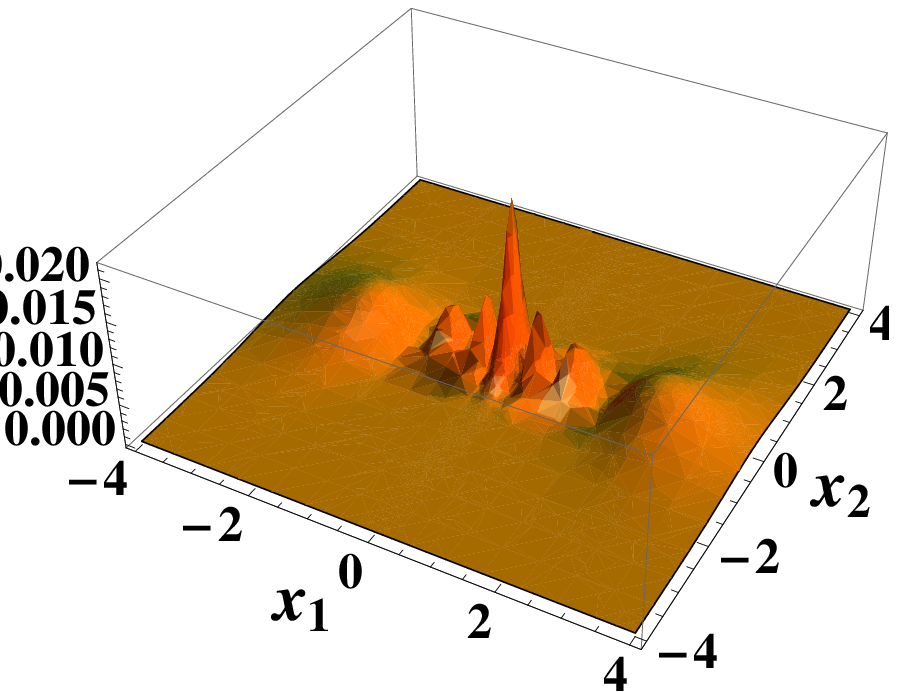}
\caption{(Color online) Exact Husimi distribution ${\Psi}^{(N)}_\xi(z_1,z_2)$ as a function of $z_{1,2}=x_{1,2}+ip_{1,2}$ 
in `momentum space' ($z_{1,2}$ imaginary; left panel) and `position 
 space' ($z_{1,2}$ real; right panel) for different values of $\xi$ (from
top to bottom: $\xi=0$, $\xi=0.3$ and $\xi=0.98$) for $N=8$. Dimensionless units.}
\label{husimifig}
\end{figure}

\begin{figure}
\begin{center}
\includegraphics[angle=-90,width=9cm]{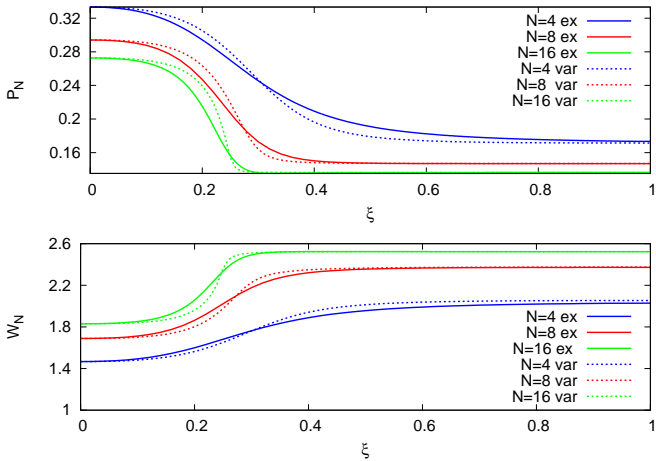}
\end{center}
\caption{(Color online)  Exact-numerical (solid) against variational-cat (dotted) inverse participation ratio 
$P_N(\xi)$ and Wehrl's entropy $W_N(\xi)$ for
  $N=4, 8$ and $N=16$ ($P_N$ decreases and $W_N$ increases with $N$)  as a function of $\xi$. Dimensionless units.}
\label{parti}
\end{figure}

Let us denote by $x_{1,2}=\mathrm{Re}(z_{1,2})$ `position' coordinates and by $p_{1,2}=\mathrm{Im}(z_{1,2})$ `momentum'  coordinates. 
In Figure \ref{husimifig} we represent the Husimi distribution in position ($p_{1,2}=0$) and momentum ($x_{1,2}=0$) cross sections. We observe 
that ${\Psi}^{(N)}_\xi(ip_1,ip_2)$ splits into two packets for $\xi\geq \xi_c=0.2$ whereas  ${\Psi}^{(N)}_\xi(x_1,x_2)$ acquires a modulation 
above the critical point $\xi_c$. We shall see below  how this `delocalization' of
the ground state is captured by {moments and Wehrl entropy of the Husimi distribution}.
We have  calculated the second moment $M_{N,2}$ (also called `inverse participation ratio' $P_N$) and the Wherl
entropy $W_N$ as a function of $\xi$. The computed results
are given in Fig. \ref{parti} (together with the variational results of Section \ref{variationalsec}), where we present $P_N(\xi)$    and $W_N(\xi)$ for
$N=4,8,16$. Notice that
the inverse participation ratio (top panel) is greater in the linear phase $\xi<0.2$ than in the bent phase $\xi>0.2$, thus capturing 
the delocalization of the Husimi distribution of the ground state across the 
critical point $\xi_c=0.2$ as depicted in Figure \ref{husimifig}. Note also that $P_N$ decreases with $N$, reaching the limiting values 
of Eq. \eqref{momentNinfty} for $\nu=2$. 
The Wherl entropy (botton panel) shows an entropy excess of $0.69\simeq \ln(2)$ [see later on Eq. \eqref{wehrlinfty}], thus capturing the splitting of the Husimi distribution 
into two non-overlapping packets in the second (bent) phase. The change of $P_N(\xi)$ and $W_N(\xi)$ across $\xi_c$ is more sudden as $N$ increases. 
Figure \ref{parti} also shows a 
good agreement between the numerical results an a variational approximation given in terms of `parity-symmetry-adapted coherent states' introduced 
in \cite{nuestro} and discussed in the next section, where we shall provide analytical 
explicit expressions for moments and Wehrl's entropy as a function of $N, \nu$ and $\xi$ and we shall discuss the thermodynamic limit 
$N\to\infty$.

\section{Variational approximation and the thermodynamic limit}\label{variationalsec}

Now we present analytical expressions for the Husimi distribution, moments and entropies using
trial states expressed in terms of ``parity-symmetry-adapted'' CSs (Schr\"odinger catlike or `cat' states for short) 
introduced by us in this context in \cite{nuestro},  
which turn out to be a good approximation to the exact (numerical)  solution of
 the ground state of the vibron model. In particular, it has been proved in  \cite{nuestro} 
  that this parity-symmetry-adapted CS captures the correct behavior
for  ground-state properties sensitive to the parity symmetry of the
Hamiltonian like  vibration-rotation entanglement and delocalization measures. 
 
\subsection{Parity-symmetry-adapted CSs and their Husimi distribution}

There are strong evidences that the exact ground state \eqref{gs} can be {\it itself} nicely approximated by a $SU(3)$ CS of the type \eqref{cohs}, when 
properly adapted to the parity symmetry (see \cite{nuestro} for more details). In fact, using the notation 
\eqref{cohs}, the particular (two-parameter) choice of `boson-condensate' denoted by
\begin{equation}
 |N;r,\theta\rangle\equiv|z_1=-\frac{r}{\sqrt{2}}e^{-i\theta}, z_2=\frac{r}{\sqrt{2}}e^{i\theta}\rangle\label{pcs}
\end{equation}
[with $r,\theta$, variational parameters representing polar coordinates] has been considered in Ref. \cite{curro} as a variational trial state to reproduce the 
ground state energy in the thermodynamic limit $N\to\infty$. Note that here $z_1$ and $z_2$ are not arbitrary complex numbers, but constrained to $z_1=-\bar z_2$. 
 As a comment, intrinsic excitations can also be constructed in this way, 
thus defining  `multi-species CSs' (see e.g \cite{kuyucak,caprio}). 

The variational parameter $r$ is fixed by minimizing the ground state energy
functional `per particle' (see \cite{curro} for more details):
\begin{eqnarray}
 {\cal E}_\xi(r)&=&\frac{\langle\hat{H}\rangle}{N}=(1-\xi)\frac{\langle\hat{n}\rangle}{N}+\xi\frac{N(N+1)-
\langle\hat{W}^2\rangle}{N(N-1)}\nonumber\\
&=&
(1-\xi)\frac{r^2}{1+r^2}+\xi
\left(\frac{1-r^2}{1+r^2}\right)^2\label{energyns}
\end{eqnarray}
where we have used here $\langle\cdot\rangle$ as a  shorthand for expectation values in $|N;r,\theta\rangle$. Note that ${\cal E}_\xi$ 
does not depend on $\theta$ because of the intrinsic rotational symmetry, so that we shall restrict ourselves to $\theta=0$ and simply write from now on 
$|N;r\rangle\equiv|N;r,\theta=0\rangle$. 
From $\partial {\cal E}_\xi(r)/\partial r=0$ one gets
the `equilibrium radius' $r_e$ and the ground state energy $ {\cal E}_\xi$
as a function of the control parameter $\xi$ (see \cite{curro} for more details):
\begin{equation}
 \begin{array}{l}
r_e(\xi)=\left\{\begin{array}{ll} 0, &  \xi\leq \xi_c=1/5\\
\sqrt{\frac{5\xi-1}{3\xi+1}}, &\xi> \xi_c=1/5 \end{array}\right.\\
 {\cal E}_\xi(r_e(\xi))=\left\{\begin{array}{ll} \xi, &  \xi\leq \xi_c=1/5\\
{\frac{-9\xi^2+10\xi-1}{16\xi}}, &\xi> \xi_c=1/5. \end{array}\right.
\end{array}\label{req1}
\end{equation}
Then one finds that $d^2{\cal E}_\xi(r_e(\xi))/d\xi^2$ is discontinuous at $\xi_c=1/5$ and the phase transition is said to
be of second order.

Although  the CS $|N;r_e(\xi)\rangle$ properly describes the mean energy 
in the thermodynamic limit $N\to\infty$ (see \cite{curro} for more details), it has been recently noticed  by us in \cite{nuestro} that it 
does not capture the correct behavior for other ground state
properties sensitive to the parity symmetry $\hat\Pi$ of the Hamiltonian like,
for instance, vibration-rotation entanglement. Here we shall see that the Husimi distribution $\Phi^{(N)}_\xi(z_1,z_2)\equiv|\langle z_1,z_2|\phi^{(N)}_\xi\rangle$ 
of $|\phi^{(N)}_\xi\rangle=|N;r_e(\xi)\rangle$ neither captures the delocalization 
of the ground state across the phase transition displayed in 
Figure \ref{husimifig} and quantified by the Wehrl entropy in Figure \ref{parti}, since it does not have a definite parity like the exact ground 
state \eqref{gs} does. Indeed, the explicit expression of the Husimi distribution of  $|\phi^{(N)}_\xi\rangle=|N;r_e(\xi)\rangle$ 
can be calculated, as a function of 
$(z_1,z_2)$, through the CS overlap \eqref{csoverlap} as:
\begin{eqnarray}
 \Phi_\xi^{(N)}(z_1,z_2)&=&|\langle z_1,z_2|\phi^{(N)}_\xi\rangle|^2=|\langle z_1,z_2|z'_1,z'_2\rangle|^2\nonumber\\
 &=&
\frac{|1- \bar z_1\frac{r_e(\xi)}{\sqrt{2}}+ \bar z_2\frac{r_e(\xi)}{\sqrt{2}}|^{2N}}{(1+|z_1|^2+|z_2|^2)^{N} (1+r_e^2(\xi))^{N}},  
\end{eqnarray}
where we have substituted $z'_1=-{r_e(\xi)}/{\sqrt{2}}, z'_2={r_e(\xi)}/{\sqrt{2}}$, as in Eq. \eqref{pcs} for $\theta=0$ and $r=r_e(\xi)$. This distribution 
$\Phi_\xi^{(N)}(z_1,z_2)$ has a single maximum at $(z_1^{(0)},z_2^{(0)})=(-{r_e(\xi)}/{\sqrt{2}},{r_e(\xi)}/{\sqrt{2}})$ and therefore does not display the two-packets structure 
of the exact distribution $\Psi_\xi^{(N)}(z_1,z_2)$ above $\xi_c$ as depicted in Figure \ref{husimifig}. We shall also see later in Section \ref{wehrlsec} 
that $\Phi_\xi^{(N)}(z_1,z_2)$ has 
constant Wehrl entropy, $W_N(\xi)={N(3+2N)}/({(N+1)(N+2)})$, and therefore it does not capture the QPT at $\xi=0.2$, where the exact Wehrl 
entropy undergoes a sudden increase, as displayed 
in Figure \ref{parti}.

The problem is that  the variational CS $|\phi^{(N)}_\xi\rangle=|N;r_e(\xi)\rangle$ does not have a definite parity, unlike the 
exact ground state $|\psi^{(N)}_\xi\rangle$, which has even-parity. A far better variational description
of the ground state is given in terms of the even-parity projected CS (see \cite{nuestro} for more details):
\begin{equation}
|N;r,+\rangle\equiv\frac{(1+\hat\Pi)|N;r\rangle}{{\cal N}_+(r)}=
\frac{|N;r\rangle+|N;-r\rangle}{{\cal N}_+(r)},\label{even-pproj}
\end{equation}
where ${\cal N}_+(r)=\sqrt{2}(1+ \langle N;-r|N;r\rangle)^{1/2}$ is a normalization constant, with  (remember the CS overlap \eqref{csoverlap})
\begin{equation}
 \langle N;-r|N;r\rangle=((1-r^2)/(1+r^2))^N.\label{overlap}\end{equation} 
Since $\langle N;-r|N;r\rangle\to 0$ when $N\to\infty$, the even-parity state \eqref{even-pproj} 
is a superposition of two weakly-overlapping (distinguishable) 
quasi-classical (coherent) wave packets, which justifies the term `Schr\"odinger catlike' for these states. Parity-symmetry-adapted CSs (of other kind) 
have also been proposed in \cite{casta1,casta2}, and used by us in \cite{epl2012,renyipra,husimidicke}, 
to study the Dicke model QPT. In particular, in \cite{husimidicke} we analyze the Husimi distribution (exact and variational) of the ground 
state in the Dicke model, which shares some features with the present vibron model.

\begin{figure}
\includegraphics[width=8.5cm,angle=0]{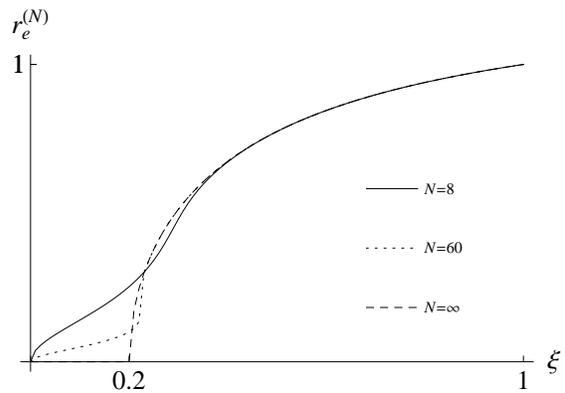}
\caption{Equilibrium radius $r_e^{(N)}(\xi)$ for $N=8,60, \infty$, where we are identifying $r_e^{(\infty)}(\xi)=r_e(\xi)$. Dimensionless units.}\label{rop-s-ns}
\end{figure}

The variational parameter $r$  in \eqref{even-pproj} is again computed by minimizing the ground state energy functional `per particle'
${\cal E}_{\xi,+}^{(N)}(r)=\langle\hat H\rangle_+/N$ as in (\ref{energyns}),
but now for the symmetric configuration \eqref{even-pproj},
given in terms of the new mean values:
\begin{eqnarray}
\frac{\langle\hat{n}\rangle_+}{N}&=&\frac{r^2((1+r^2)^{N-1}-(1-r^2)^{N-1}}{(1+r^2)^{N}+(1-r^2)^{N}},\\
\frac{\langle\hat{W}^2\rangle_+}{N}&=&2\frac{(1+r^2)^{N}+(1-r^2)^{N-2}(1+2Nr^2+r^4)}{(1+r^2)^{N}+(1-r^2)^{N}}.\nonumber
\end{eqnarray}
Unlike ${\cal E}_\xi(r)$,  the new energy functional ${\cal E}_{\xi,+}^{(N)}(r)$ depends on $N$.
From $\partial {\cal E}_{\xi,+}^{(N)}(r)/\partial r=0$ we can obtain the new equilibrium radius $r_e^{(N)}(\xi)$. We do not have an explicit 
analytic expression of $r_e^{(N)}(\xi)$ (as we did for $r_e(\xi)$ in \eqref{req1}) for arbitrary $N$ and $\xi$, but one can always compute it numerically. 
Figure \ref{rop-s-ns} compares  $r_e(\xi)$ in (\ref{req1}) with $r_e^{(N)}(\xi)$  for
$N=8$ and $N=60$. We can infer that, in the thermodynamic limit,  $r_e^{(\infty)}(\xi)=r_e(\xi)$. This is a curious fact. 

The Husimi distribution of the (even) cat state $|\phi^{(N)}_{\xi,+}\rangle\equiv|N;r_e^{(N)}(\xi),+\rangle$ in \eqref{even-pproj} is then
\begin{eqnarray}
 \Phi_{\xi,+}^{(N)}(z_1,z_2)&=&|\langle z_1,z_2|\phi^{(N)}_{\xi,+}\rangle|^2\nonumber\\
 &=& \frac{|\langle z_1,z_2|N;r\rangle+\langle z_1,z_2|N;-r\rangle|^2}{{\cal N}_+^2(r)},\label{husvar}
\end{eqnarray}
with
\begin{equation}
\langle z_1,z_2|N;\pm r\rangle=
\frac{(1\mp \frac{r}{\sqrt{2}}\bar z_1\pm \frac{r}{\sqrt{2}}\bar z_2)^N}{(1+|z_1|^2+|z_2|^2)^{N/2} (1+r^2)^{N/2}},\label{csoverlap2}
\end{equation}
where we must understand now $r=r_e^{(N)}(\xi)$. Since $|\langle z_1,z_2|N;\pm r\rangle|$ is maximum at 
$(z_{1,\pm}^{(0)},z_{2,\pm}^{(0)})=(\mp\frac{r}{\sqrt{2}},\pm\frac{r}{\sqrt{2}})$, the variational 
Husimi distribution \eqref{husvar} captures the exact two-packets structure displayed in Figure \ref{husimifig}, although rotated $\pi/4$. 
Moments and Wehrl's entropy are insensitive to this global rotation, so that we keep this variational state for which explicit 
expressions and calculations turn out to be simpler.

\subsection{Moments and Wehrl-Lieb conjecture}\label{wehrlsec}

Explicit analytic expressions for the $\nu$-th moment $M_{N,\nu}^+$ of the variational-cat Husimi distribution \eqref{husvar} are 
easily obtained for the rigidly linear phase $\xi=0$ ($r=r_e^{(N)}(0)=0$) giving:
\begin{eqnarray}
M_{N,\nu}^+(0)&=& \int_{{\mathbb R}^4}({\Phi}^{(N)}_{0,+}(z_1,z_2))^\nu  d\mu(z_1,z_2)\nonumber\\ &=&\frac{(N+1)(N+2)}{(1+\nu N)(2+\nu N)},\label{momentvar}
\end{eqnarray}
which are in complete agreement with exact-numerical calculations, as displayed in Figure \ref{parti} for the particular case of $\nu=2$.  
The same happens to the Wehrl entropy for $\xi=0$, which can be obtained as:
\begin{equation}
W_N^+(0)=\lim_{\nu\to 1}\frac{1}{1-\nu}\ln M_{N,\nu}^+(0)=\frac{N(3+2N)}{(N+1)(N+2)}.\label{wehrlvar}
\end{equation}
For other values of $\xi$, integrals can always be  numerically done (see e.g. Figure \ref{parti} for $M_{N,2}(\xi)$ and 
$W_N(\xi)$ as a function of $\xi$ for $N=4,8,16$). In the rigidly bent phase $\xi=1$, we have been able to obtain the asymptotic behavior 
for $N\to\infty$, in particular:
\begin{equation}
M_{N,\nu}^+(\xi)\stackrel{N\to\infty}{\longrightarrow}\left\{\begin{array}{ll} {\nu^{-2}}, & \mathrm{if}\, \xi=0\\
{2^{1-\nu}\nu^{-2}}, &
\mathrm{if}\, \xi=1,\end{array}\right.\label{momentNinfty}
\end{equation}
and
\begin{equation}
W_N^+(\xi)\stackrel{N\to \infty}{\longrightarrow}\left\{\begin{array}{ll} 2, & \mathrm{if}\, \xi=0\\
2+\ln(2), &
\mathrm{if}\, \xi=1.\end{array}\right.\label{wehrlinfty}
\end{equation}
In fact, expressions \eqref{momentNinfty} and \eqref{wehrlinfty} also give a good approximation for $N\gg 1$ in the `floppy' region 
$0<\xi<1$ since changes in moments and Wehrl entropy are sharper and sharper as $N$ increases, both being approximately 
constant in each phase as inferred in Figure \ref{parti}.

At this point, one could ask himself/herself to what extent are the results for the parity-symmetry-adapted CS \eqref{even-pproj} better or different from the 
ordinary CS ansatz \eqref{pcs}.  We must say that Wehrl entropy for the Husimi distribution $\Phi_\xi^{(N)}$ of the (ordinary) CS \eqref{pcs} 
is {\it constant}, $W_N(\xi)={N(3+2N)}/({(N+1)(N+2)})$, as a 
function of the control parameter $\xi$ (through the dependence of $r=r_e(\xi)$) and, therefore, it does not capture the delocalization 
of the ground state across the phase transition displayed in  
Figure \ref{husimifig} and quantified by the IPR and Wehrl entropy in Figure \ref{parti}.  
On the contrary, the Husimi distribution $\Phi_{\xi,+}^{(N)}$ in Eq. \eqref{husvar} nicely captures this delocalization, exhibiting an entropy excess 
of $\ln(2)$ from linear to bent phases, in agreement with exact numerical results.

We should also point out that the behavior displayed in \eqref{momentNinfty} and \eqref{wehrlinfty} has also been found by us in the Dicke model 
of matter-field interactions in the thermodynamic limit $N=2j\to\infty$ (the number of atoms), which also exhibits a QPT from normal 
to superradiant (see \cite{husimidicke}).

To finish this section, we would like to comment on the still unproved Lieb's conjecture. It was conjectured by Wehrl \cite{Wehrl} 
and proved by Lieb \cite{Lieb} that any Glauber (harmonic oscillator) coherent state $|\alpha\rangle$ has a minimum Wehrl entropy of 1. 
In the same paper by Lieb \cite{Lieb}, it was also conjectured that the extension of
Wehrl's definition of entropy for spin-$j$ CSs \eqref{spinjcs} 
will yield a minimum entropy of $2j/(2j+1)$. Here we propose that the extension \eqref{wehrl1} of 
Lieb's definition of entropy  will yield a minimum entropy of ${N(3+2N)}/({(N+1)(N+2)})$ for $SU(3)$ projective 
CSs \eqref{cohs}. We have seen that the ground state of the 
vibron model in the rigidly linear phase ($\xi=0$) is itself a $SU(3)$ CS and its Wehrl entropy reaches this minimum. 
The same value is attained for the Wehrl entropy of any other $SU(3)$ CS like \eqref{cohs}. In the rigidly bent phase ($\xi=1$), the 
ground state is not a $SU(3)$ CS anymore, but has a ``cat-like'' structure (linear combination of CSs) giving a Wehrl entropy excess of 
$\ln(2)$.

\subsection{Zeros of the variational Husimi distribution}

It is well known that the Husimi density is determined by its zeros through the Weierstrass-Hadamard factorization. 
It has also been observed that the distribution of zeros differs for classically regular or
chaotic systems and can be considered as a quantum indicator of classical
chaos (see e.g. \cite{Korsch,monteiro,4}).   Moreover, recently we have
presented  a characterization
of the Dicke model QPT by means of the zeros of the Husimi distribution in the variational
approach \cite{husimidicke}.

Here we shall explore the distribution of zeros of the Husimi density as a fingerprint 
of QPT in the vibon model. From \eqref{husvar} we obtain
\begin{equation}
 {\Phi}^{(N)}_{\xi,+}(z_1,z_2)=0\Rightarrow  z_2-z_1=\frac{i\sqrt{2}}{r^{(N)}_e(\xi)}\tan(\frac{(2l+1)\pi}{2N})
\end{equation}
with $l=-[N/2],\dots,[N/2]-1$ and $[N/2]=$Floor$(N/2)$. If we separate real an imaginary parts as 
$z_{1,2}=x_{1,2}+ip_{1,2}$, the last condition can be cast as
\begin{eqnarray}
 x_1&=&x_2,\\
 p_2&=&p_1+\frac{\sqrt{2}}{r^{(N)}_e(\xi)}\tan(\frac{(2l+1)\pi}{2N}).\label{zeros2}
\end{eqnarray}
For $r^{(N)}_e(\xi)=0$ the Husimi distribution $\Phi^{(N)}_{\xi,+}(z_1,z_2)$ has no zeros. For finite $N$, the value $r^{(N)}_e(\xi)=0$ 
is only attained in the rigidly linear phase $\xi=0$ (see Figure \ref{rop-s-ns}). In the thermodynamic limit $N\to\infty$ we have 
that $r^{(\infty)}_e(\xi)=r_e(\xi)=0, \forall \xi<\xi_c=0.2$ (See Figure \ref{rop-s-ns} and expression \eqref{req1}) so that ${\Phi}^{(\infty)}_{\xi,+}(z_1,z_2)$ 
has no zeros in the linear phase. For $r^{(N)}_e(\xi)\not=0$
the zeros are localized along straight lines (``dark fringes'') in the 
$x_1x_2$ (position) and $p_1p_2$ (momentum) planes. In the momentum plane, the density of zeros grows with $N$ and $\xi$ 
(viz, with $r^{(N)}_e(\xi)$, since it is an increasing function of $xi$). In the 
thermodynamic limit $N\to\infty$, there is a sudden growth of zeros for $\xi>\xi_c=0.2$ which accumulate in a vicinity of 
$p_2=p_1$ in the momentum plane $p_1p_2$. 
A similar behavior is also shared by the Dicke model (see \cite{husimidicke}).

\section{Conclusions}\label{sec3}

We have found that moments and Wehrl entropies of the Husimi distribution provide 
sharp indicators of a quantum phase transition in the vibron model. They detect a delocalization of the 
Husimi distribution across the critical point $\xi_c$ and we have employed them to quantify the phase-space spreading of
the ground state.

Calculations have been done numerically and through a variational
approximation.
{We have represented the Husimi distribution which exhibits a
  different shape in each phase. We have calculated the inverse participation
  ratio and the Wehrl entropy, which prove to be  good indicators of  the QPT. }
 The variational approach, in terms of parity-symmetry-adapted 
coherent (cat) states, complements and 
enriches the analysis  providing explicit analytical
expressions for the moments and Wehrl entropies which remarkably coincide with the
numerical results, especially in the rigidly linear and bent phases (outside the floppy region $\xi\approx\xi_c$) and 
in the thermodynamic limit.

In the rigidly linear phase, Wehrl's entropy attains its minimum $\frac{N(3+2N)}{(N+1)(N+2)}$, according to a generalized 
Wehrl-Lieb conjecture, thus indicating that the ground state for $\xi=0$ is a $SU(3)$ projective CS. 
In the bent phase, 
Wehrl's entropy undergoes an 
entropy excess (or ``subentropy'' \cite{Jozsa}) of $\ln(2)$. This fact implies that the Husimi distribution splits 
up into two identical subpackets with negligible overlap in passing from linear to bent phase; in general, for 
$s$ identical subpackets with negligible
overlap, one would expect an entropy excess of $\ln(s)$. This delocalization of the exact Husimi distribution in the bent phase 
is not captured by the variational approximation in terms of the ordinary $SU(3)$ CS \eqref{pcs}, which gives a constant value of IPR $P_N$ and Wehrl entropy 
$W_N$ across the phase transition.   On the contrary, the parity-symmetry-adapted CS of eq. \eqref{even-pproj} nicely reproduces the 
exact ground state behavior, as seen in figure \ref{parti}.

The QPT fingerprints in the vibron model have also been tracked by exploring
the distribution of zeros of the Husimi density within the
  analytical variational approximation. We have found that there is a sudden growth of zeros above the 
  critical point $\xi_c$, specially in the thermodynamic limit. Zeros of the variational Husimi (cat) distribution 
exhibit a richer structure in momentum than in position space. 
This behavior is also shared by the Dicke model \cite{husimidicke}. 
 This subject 
deserves further attention and will be studied in future works. 

The different structure of zeros of 
the Husimi distribution for classically regular or
chaotic systems has also been considered in the literature as a quantum indicator of classical chaos (see e.g. \cite{Korsch,monteiro,4}). 
For example, in \cite{4} it is shown that, in integrable regions, the zeros lie on one-dimensional curves, 
while in chaotic regions the distribution is bi-dimensional and the zeros
fill the phase-space. 
 We have restricted ourselves to the phase-space analysis
of the ground state in the vibron-model. {The vibron-model is
  a regular (non chaotic) system  so we can assert that 
the sudden growth of zeros above the critical point $\xi_c$ denotes a QPT but it is not 
a symptom of chaos.}

\section*{Acknowledgments}

This work was supported by the Projects:   FIS2011-24149 and FIS2011-29813-C02-01 (Spanish MICINN),  
FQM-165/0207 and FQM219 (Junta de Andaluc\'\i a) and 20F12.41 (CEI BioTic UGR).

\end{document}